\documentclass[aps,prl,twocolumn,groupedaddress]{revtex4-1}

\usepackage{graphicx}
\usepackage{dcolumn}
\usepackage{bm}

\begin{document}


\title{Beam Blow Up due to Beamstrahlung in Circular e$^+$e$^-$ Colliders}


\author{M.A.~Valdivia Garcia}
\email{valdivia@fisica.ugto.mx}
\affiliation{U.~Guanajuato, 36000 Guanajuato, 
Mexico}

\author{F.~Zimmermann}
\affiliation{CERN, 1211 Geneva 23, Switzerland }



\date{\today}

\begin{abstract}
After the discovery of the Higgs boson at the Large Hadron Collider in 2012, 
several possible 
future circular colliders --- Higgs factories --- 
are proposed, 
such as FCC-ee and CEPC. 
At these highest-energy  e$^+$e$^-$  
colliders, beamstrahlung, namely 
the synchrotron radiation emitted in the field of the opposing beam, 
can greatly affect the equilibrium bunch length and energy spread. If the  dispersion function at the collision point is not zero, beamstrahlung will also increase 
the transverse emittances. 
In this letter, we first show that, for circular Higgs factories, a classical description of the beamstrahlung is adequate. 
We then derive analytical formulae describing the equilibrium  beam parameters, taking into account the
variation of the electromagnetic field during the collision. We illustrate the importance of beamstrahlung, including the increase of bunch length and the implied tolerance on the spurious dispersion function at the collision point, by considering a few examples.

%
\end{abstract}

\pacs{}

\maketitle

In most electron storage rings operated so far, 
the equilibrium transverse 
emittances, energy spread and bunch length  
were, or are, determined by a balance of quantum excitation and radiation damping, both arising from the 
synchrotron radiation
emitted when the charged ultra-relativistic beam particles pass through the accelerator magnets, in particular through the bending magnets  
\cite{Sands1970ye}.  
Future high-energy circular colliders, like FCC-ee \cite{fccee} or CEPC \cite{cepc}, are 
proposed as high-precision Higgs factories,
to study the Higgs boson discovered at the Large Hadron Collider \cite{higgs},
or, more generally, as 
``electroweak factories''.
In these future circular colliders, for the first time, 
also the synchrotron radiation emitted during the collision 
in the electromagnetic field of the opposing beam 
becomes important. This particular type of 
synchrotron radiation is called
 ``beamstrahlung''   \cite{beams1,beams2,beams3,Blankenbecler1987rg,Bell1987ve,yokoya,yokoyachen}. 
A beam particle is lost whenever, during the collision, it 
radiates a photon of an energy high enough  
that the emittance particle falls outside the 
momentum acceptance. 
Through this process,  
the high-energy tail of the
can severely limit the beam lifetime  
\cite{telnov,bogomprstab}. 
Design parameters for FCC-ee and CEPC are 
taking into account this lifetime limitation
along with additional 
constraints imposed by a 
coherent beam-beam instability 
\cite{Ohmi2017cwi}.  

There is yet another novel effect of beamstrahlung in circular Higgs factories.   
Namely, at the aforementioned colliders 
the beamstrahlung 
significantly increases the 
equilibrium bunch length and energy spread
of the colliding beams \cite{yokoya2,ohmifz,ipac16bs}.  
Furthermore, with a non-zero dispersion at the IP, beamstrahlung can also 
affect the transverse beam emittance \cite{ipac16bs,ipac16mc}. 
Such nonzero dispersion can either be due to 
incompletely corrected optics errors  (``spurious dispersion'') 
or be intentionally introduced for the purpose of
reducing the centre-of-mass energy spread
(``monochromatization'') \cite{renieri}.

The strength of the synchrotron radiation is characterized by the
parameter $\Upsilon$, defined  as \cite{yokoya,yokoyachen}
$\Upsilon\equiv B/B_{c} = (2/3) \hbar \omega_{c}/E_{e}$, with 
$B_{c}=m_{e}^{2}c^{2}/(e\hbar)\approx 4.4$~GT the Schwinger critical field,
$\omega_{c}$ the critical photon energy as 
defined by Sands \cite{Sands1970ye}, 
and $E_{e}$ the electron energy before radiation.

For the collision of 3-dimensional Gaussian 
bunches with rms sizes $\sigma_{x}^{\ast}$, $\sigma_{y}^{\ast}$ and $\sigma_{z}$,
possibly under a small horizontal 
crossing angle $\theta_{c}$, 
the average $\Upsilon$ is 
\cite{yokoyachen}
\begin{equation}
\left\langle \Upsilon \right\rangle  \approx \frac{5}{6}
\frac{r_{e}^2 \gamma N_{b}}{\alpha \sigma_z (\sigma_x^\ast+\sigma_y^{\ast} )}  \;  ,  
\end{equation}
where $\alpha$ denotes 
the fine structure constant ($\alpha\approx 1/137$),
and $r_{e}\approx 2.8\times 10^{-15}$~m the classical electron radius.

For all proposed high-energy circular e$^+$e$^-$ colliders, 
$\Upsilon \ll 1$ and $\sigma_{x}^{\ast}\gg \sigma_{y}^{\ast}$. 
In this case we can approximate   
the average number of photons per collision
as \cite{yokoyachen,telnov,bogomprstab} 
\begin{equation}
 n_\gamma 
\approx
\frac{12}{\pi^{3/2}} 
\frac{\alpha r_e  N_b}{\sigma_x^{\ast}}
\frac{1}{\sqrt{1+\Phi_{\rm piw}^{2}}}\; , 
\label{ngamma}
\end{equation}
where $\Phi_{\rm piw}\equiv \theta_{c} \sigma_{z}/(2 \sigma_{x}^{\ast})$ is a geometric reduction factor,
also known as the ``Piwinski angle''.  
The average relative energy loss is \cite{yokoyachen} 
\begin{equation}
\delta_B 
\approx
\frac{24}{3\sqrt{3}\pi^{3/2}}\, 
\frac{r_e^3 \gamma N_b^2}{\sigma_z {\sigma_{x}^{\ast}}^2 }
\frac{1}{\sqrt{1+\Phi_{\rm piw}^{2}}}  
\; .
\end{equation}

The average photon energy normalized to the beam energy, $<u>$, 
is given by the ratio of $\delta_{B}$ and $n_{\gamma}$:
\begin{equation}
\langle u \rangle = \frac{\delta _{B}}{n_{\gamma}} \approx 
\frac{2\sqrt{3}}{9} \frac{r_{e}^{2}N_{b}\gamma}{\alpha \sigma_{z}
\sigma_{x}^{\ast} }\; .
\end{equation}

The quantum excitation, which gives rise to energy spread and emittance, 
is the product of the mean square photon energy and the mean emission 
rate \cite{Sands1970ye}.
In the case of beamstrahlung, 
the mean rate is simply given by $n_{\gamma}$ divided by the average time interval between collisions, e.g.,
half the revolution period in case of two interaction points. 
Introducing $y \equiv \omega / E _ { e }$ and
\begin{equation}
    \xi \equiv \frac { 2 \omega } { 3 \Upsilon ( E - \hbar \omega ) }\; , 
\end{equation}
the emission rate spectrum (photons emitted per second per
energy interval) is described by the function
\cite{yokoyachen,sokolovternov} 
\begin{equation}
\frac { d W _ { \gamma } } {\hbar d \omega } = \frac { \alpha } { \sqrt { 3 } \hbar \pi \gamma ^ { 2 } } \left( \int _ { \xi } ^ { \infty } K _ { 5 / 3 } \left( \xi ^ { \prime } \right) d \xi ^ { \prime } + \frac { y ^ { 2 } } { 1 - y } K _ { 2 / 3 } ( \xi ) \right)  \; ,  
\label{Wgen}
\end{equation}
which in the classical regime $( \Upsilon \rightarrow 0 )$ reduces to
\cite{Sands1970ye} 
\begin{equation}
    \frac { d W _ { \gamma } } {\hbar d \omega } = \frac { \alpha } { \sqrt { 3 } \pi \gamma ^ { 2 } } \int _ { \xi } ^ { \infty } K _ { 5 / 3 } \left( \xi ^ { \prime } \right) d \xi ^ { \prime }\; .
    \label{Wclass}
\end{equation}
The number of photons radiated per unit time is obtained
by integrating over $\omega$:
\begin{equation}
    \frac { d N _ { \gamma } } { d t } = \int _ { 0 } ^ { E _ { e } / \hbar } \frac { d W _ { \gamma } } { d \omega } d \omega
\end{equation}

\begin{figure}[htb]
    \centering
    \includegraphics*[width=0.9\columnwidth]{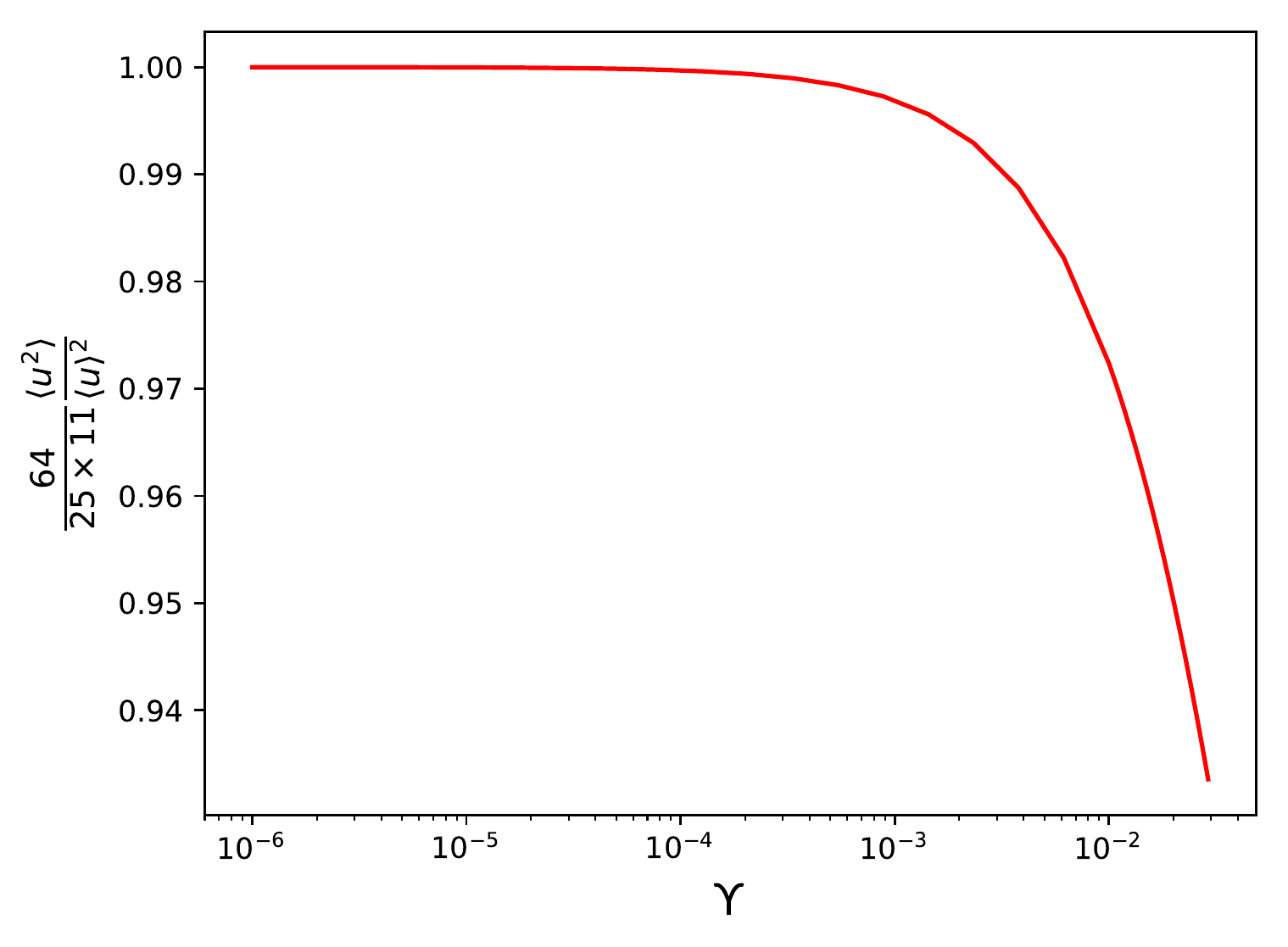}
    \caption{Mean square photon energy normalized by the square of the mean energy according to (\ref{u2exact}), for a constant bending radius $\rho$. The vertical scale is normalized such that the approximate relation 
    (\ref{u2}) corresponds to a value of 1.}
    \label{fig1}
\end{figure}

In the classical radiation regime and for a constant bending radius $\rho$, 
the mean square photon energy $<u^{2}>$ is related to the average
photon energy $<u>$ via \cite{Sands1970ye}
\begin{equation}
\langle  u^{2} \rangle   \approx  
 \frac{25 \times 11}{64} \langle u \rangle ^{2}\; \; \; \; {\rm (constant\; }\rho{\rm )}\; ,
\label{u2}
\end{equation}
where $<u>\propto 1/\rho$. 
Using the function 
$dW_{\gamma} /d\omega$ of (\ref{Wgen}),
we can numerically determine the 
exact ratio 
\begin{equation}
\frac{<u^2>}{<u>^2} = \frac{dN_\gamma}{dt} \frac{\int _{0}^{E_{e}/\hbar} (d W _ { \gamma } \omega^2/ d \omega) d\omega }{(\int _{0}^{E_{e}/\hbar} (d W _ { \gamma } \omega/ d \omega) d\omega )^2 } 
\label{u2exact} 
\end{equation}
for a constant value of the 
critical photon energy or of the bending radius. 
The result, shown in Fig.~\ref{fig1}, 
demonstrates that the error of the 
classical relation (\ref{u2})
is smaller than 1\% 
for $\Upsilon$ values 
up to several times $10^{-3}$
~\cite{ipac16mc,cremlin,ipac17}.

The classical formulae for synchrotron radiation 
would also be modified for an interaction length
($\approx \sigma_{x}^{\ast}/\theta_{c}$)
shorter than the classical formation 
length $\rho/\gamma$ \cite{telnov18,coisson},
with $\rho$ the local bending radius.
This ``short-magnet''  regime
is characterized by an ``undulator  parameter'' $K_{\rm max} \equiv 
2 r_{e} N_{b}/(\sigma_{x}\theta_{c})< 1$,
while the classical radiation 
spectrum applies for $K_{\rm max}> 1$.
As we will see below, for all the cases of 
interest $K_{\rm max}\ge 3$, so that the effect of 
short-magnet radiation can be neglected.

In the case of a real bunch collision,
the relation between $<u>$ and $<u^{2}>$ 
is further modified, however, for another reason:  
The local bending 
radius is not constant, 
but varies with the transverse and longitudinal position of the colliding particle,
and with the time during the collision \cite{oidepriv,ipac18}.  
Indeed, while at constant bending radius $\rho$ 
we have \cite{Sands1970ye}
$ \langle u \rangle = 4/(5 \sqrt{3}) \hbar c \gamma^{3} / \rho $,  
and $<u^2>$ well represented by (\ref{u2}),      
in general (\ref{u2}) must be modified as 
\begin{equation}
\langle u^{2} \rangle   \approx Z_{\rm c} 
 \frac{25 \times 11}{64} \langle u \rangle^{2}\; 
\label{u2cor}
\end{equation}
where the correction factor $Z_{\rm c}$ is related to the  
variation of $1/\rho$ in time and space:
\begin{equation}
Z_{\rm c} \equiv \frac{\left\langle 1/\rho^{2} \right\rangle_{x,y,s,t}}{ \langle 1/ \rho \rangle _{x,y,s,t} ^{2} }  \; ,
\label{zcor}
\end{equation}
where $\langle ...\rangle $
denotes the bunch average  
in space $(x,y,s)$ and time $t$ during a collision. 


To treat the case of a nonzero  
crossing angle, we consider the
collision in a co-moving (boosted) frame
\cite{hirata},
where the collision is ``head-on'',
but both bunches are tilted by 
an opposite angle of magnitude
$\theta_{c}/2$. 
In first approximation, 
for the future circular colliders considered, 
we may ignore the
disruption effects \cite{yokoyachen},
and we also neglect the
rms angular beam divergence 
compared with the
crossing angle $\theta_{c}$.
We do take into account
the vertical hourglass effect by
considering a vertical rms
beam size which changes
with longitudinal position $s$ as
\begin{equation}
\sigma_{y}(s) = \sqrt{\varepsilon_{y} \left( \beta_{y}^{\ast} + \frac{s^2}{\beta_{y}^{\ast}}\right) }\; ,  
\end{equation}
where $\varepsilon_{y}$ denotes the vertical rms
emittance and $\beta_{y}^{\ast}$ the 
vertical beta function at the focal point. 
Under these assumptions, 
the inverse local bending radius 
$\rho$ at transverse 
coordinates ($x$, $y$), and longitudinal coordinates $s$ (along the beam line)
and $z$ (co-moving, along the bunch, with $z=0$ referring to 
the centre of the bunch, and $z=(s-ct)$; where $t$ is time and $c$ the speed of light)  
can be approximated as \cite{bbdb} 

\begin{eqnarray}
\lefteqn{
\frac{1}{ \rho (x,y,s,z ) } = \left|
\mathcal{F} 
\left( 
x-s\frac{\theta_c}{2},y, \sigma_{y}(s) \right)
\right| 
} 
\nonumber
\\
& & 
\frac{2 N_b r_e}{\gamma \sigma_{z}}
\sqrt{\frac{2}{\pi }} 
\exp \left( 
- \frac{2 (s-\frac{z}{2})^2}{\sigma_{z} ^2}
\right)
\label{rho}
\end{eqnarray}
where $ \mathcal{F} (x,y,\sigma_{y}(s))$ 
may be expressed 
%
in terms of
the complex error function $w$ as
\cite{basers} 
\begin{eqnarray}
\lefteqn{\mathcal{F} (x, y,\sigma_y(s))
=
\frac{\sqrt{\pi}}{\sqrt{2}( {\sigma_{x}}^{2} -{\sigma_{y}}^{2}(s))} 
}
\nonumber
\\
& & 
\left(
w
\left[
\frac{x + iy}{\sqrt{2({\sigma_{x}}^{2} -{\sigma_{y}}^{2}(s))}}
\right]  \right.
\nonumber 
\\ & & 
\left. 
-
e^{
-\frac{x^{2}}{{2\sigma_{x}}^{2}} -  \frac{y^{2}}{{2\sigma_{y}}^{2}(s)}
}
w
\left[
\frac{ \frac{x\sigma_{y}(s)}{\sigma_{x}} + i\frac{y\sigma_{x}}{\sigma_{y}(s)}}
{
\sqrt{2({\sigma_{x}}^{2} -{\sigma_{y}}^{2}(s) )} 
}
\right] 
\right)\; .
\label{Fcomp} 
\end{eqnarray}

Including the crossing angle
and the hourglass effect,
the average inverse bending radius is
obtained as a quadruple
integral of (\ref{rho}) over 
the four dimensions:
\begin{eqnarray}
\lefteqn{
\left< 
\frac{1}{\rho }
\right> 
= \int_{x,y,z,s}
dx\; dy \; dz\; ds\;
\frac{1}{\rho (x,y,s,z)  }
}
\nonumber \\
& & 
\frac{
\exp \left(
-\frac{
\left(x+z\frac{\theta_{c}}{2}\right)^2}{2\sigma_{x}^{2}}
-\frac{y^2}{2\sigma_{y}^{2}(s)}
-\frac{z^2}{2\sigma_{z}^{2}}
-\frac{2 \left(s-\frac{z}{2}\right)^2}{\sigma_z^2} 
\right)
}
{2 \pi^{2}\sigma_x \sigma_y(s) \sigma_z^2}\; ,
\label{invrho}
\end{eqnarray}
which can be evaluated numerically. 
Similarly, we write
\begin{eqnarray}
\lefteqn{
\left< 
\frac{1}{\rho ^2}
\right> 
= \int_{x,y,z,s}
dx\; dy \; dz\; ds\;
\frac{1}{\rho (x,y,s,z)^2}
}
\nonumber \\
& & 
\frac{
\exp \left(
-\frac{
\left(x+z\frac{\theta_{c}}{2}\right)^2}{2\sigma_{x}^{2}}
-\frac{y^2}{2\sigma_{y}(s)^{2}}
-\frac{z^2}{2\sigma_{z}^{2}}
-\frac{2 \left(s-\frac{z}{2}\right)^2}{\sigma_z^2}
\right)
}
{2 \pi^2
\sigma_x \sigma_y(s) \sigma_z^2
}\; .
\label{invrho2}
\end{eqnarray}

\begin{figure}[htb]
    \centering
    \includegraphics*[width=0.9\columnwidth]{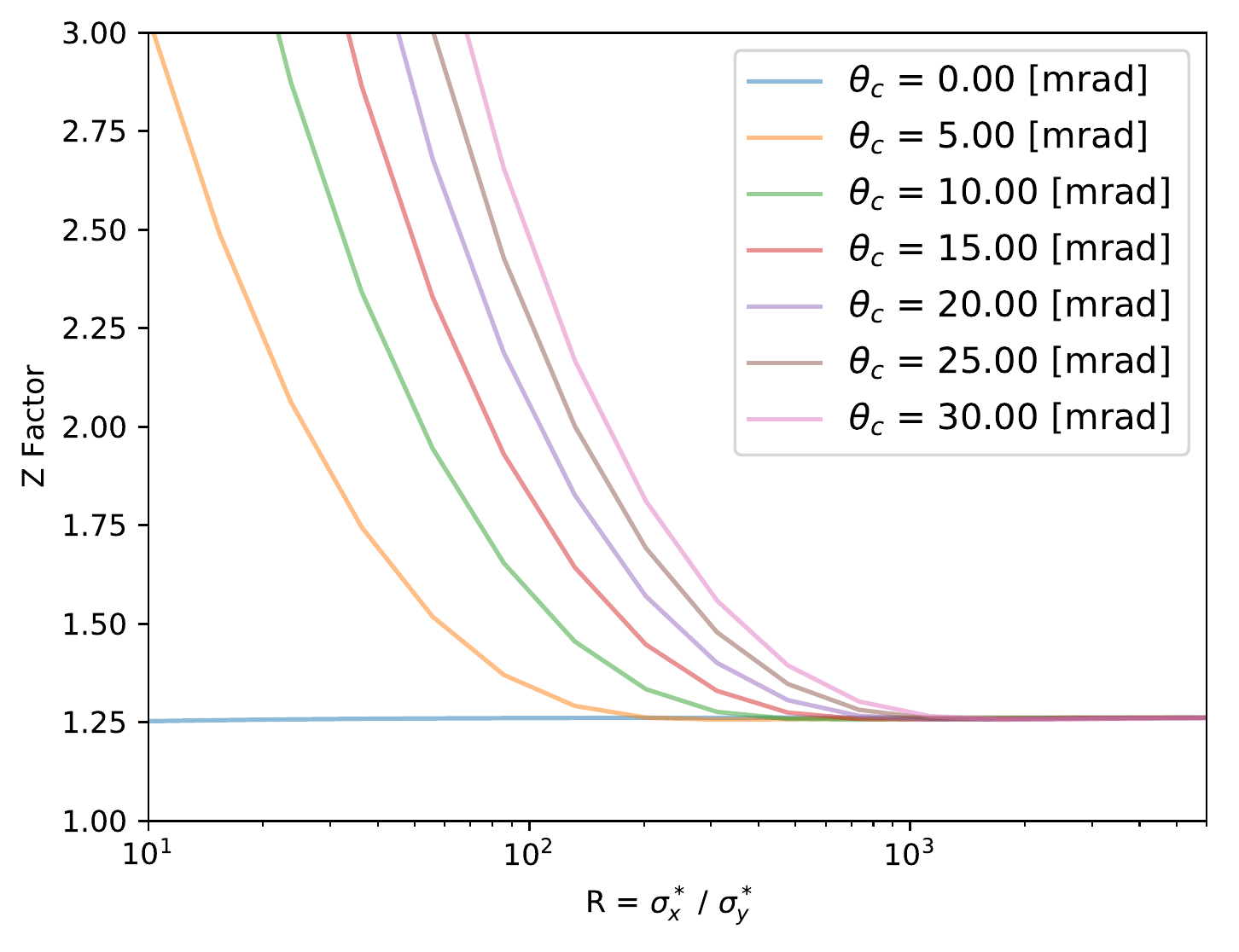}
    \caption{Correction 
factor $Z_{\rm c}$ versus $R=\sigma_{x}^{\ast}/\sigma_{y}^{\ast}$,
for different values of the full crossing angle (colors), with
 $\beta_{y}^{\ast}=1$~mm,  $\sigma_{y}^{\ast} = 32$~nm,  and 
 $\sigma_{z} = 1.64$~mm held constant.}
    \label{fig1a}
\end{figure}

Using Eqs.~(\ref{invrho}), and (\ref{invrho2}) we 
compute the correction factor $Z_{\rm c}$,
which is illustrated in Fig.~\ref{fig1a} as a 
function of the transverse beam-size aspect ratio at the collision point, for different
values of crossing angle, holding the beta function $\beta_{y}^{\ast}$~mm, 
the vertical rms beam size $\sigma_{y}^{\ast}$,  and 
the bunch length $\sigma_{z}$ constant.

Putting everything together, the quantum
excitation \cite{Sands1970ye} 
from beamstrahlung
emitted in a single collision
can be written 
\begin{equation}
   \{  n _ { \gamma } \left\langle u ^ { 2 }  \right\rangle 
   \} _{\rm BS} 
   \approx 
   \frac{25\times 11}{36 \pi^{3/2}} 
   Z_{c}   
   \frac{r_{e}^{5} N_{b}^{3}
   \gamma^{2}} {\alpha \sigma_{z}^{2} 
   {\sigma_{x}^{\ast}}^{3}} 
   \frac{1}{\sqrt{1+\Phi_{\rm piw}^{2}}} 
   \; .
\end{equation}

Balancing the sum of the excitation due to beamstrahlung and due to arc synchrotron radiation
against the radiation damping from the arcs 
alone (the average energy loss and, hence, 
the damping effect due to beamstrahlung 
is negligible \cite{ipac16bs}) yields the 
total equilibrium emittance $\varepsilon_{x,{\rm tot}}$ 
\begin{equation}
    \varepsilon _ {x,{\rm tot }} = 
    \varepsilon _ {x, {\rm SR}} + \frac {n_{\rm IP} \tau_{x,{\rm SR}} } { 4 T _ {\rm rev}} \left\{ n _ {\gamma } < u ^ { 2 } > \right\}_{\rm BS} \mathcal{H} _ { x } ^ { * }\; ,
\end{equation}
and relative rms momentum spread $\delta_{\rm tot}$ as  
\begin{equation}
    \sigma _ { \delta , {\rm tot}} ^ { 2 } = \sigma _ { \delta , {\rm SR} } ^ { 2 } + \frac { n _ { \mathrm { IP } } \tau _ { E  , \mathrm { SR } }} { 4 T _ {\rm  rev  } } \left\{ n _ { \gamma } < u ^ { 2 } > \right\}_{\rm BS} \;,
\end{equation}
where $\tau_{x}$ and $\tau_{E}$ denote the usual horizontal and longitudinal radiation
damping times \cite{Sands1970ye}, respectively, $T_{\rm rev}$ the revolution period, 
and $n_{IP}$ the number of interaction points.     
The terms with subindex ${\rm SR}$ refer to the
standard equilibrium parameters without
beamstrahlung.
The dispersion invariant 
${\cal H}_{x}^{\ast} $ is defined as \cite{Sands1970ye}  
\begin{equation}
{\cal H}_{x}^{\ast} \equiv 
\frac{\left (\beta_{x}^{\ast} {D_{x}'}^{\ast} + \alpha_{x}^{\ast}
D_{x}^{\ast}\right)^2 
+ {D_{x}^{\ast}}^{2}}{\beta_{x}^{\ast}} \; , 
\end{equation}
where $\beta_{x}^{\ast}$, $\alpha_{x}^{\ast}$, $D_{x}^{\ast}$ 
and ${D_{x}'}^{\ast}$ denote optical beta and alpha 
function (Twiss parameters), the dispersion and slope of the 
dispersion at the collision point, respectively.

The beamstrahlung parameters ($\Upsilon$, 
$\delta_{B}$, $<u>$ and $\rho$) 
strongly depend on the bunch length.
The ``total'' (equilibrium) bunch length 
is related to the total  energy spread 
via the classical relation \cite{Sands1970ye} 
\begin{equation}
\sigma_{z,{\rm tot}}= \frac{\alpha_{\rm C} C}{2\pi Q_{s}} 
\sigma_{\delta,{\rm tot}}\; , 
\label{sz}
\end{equation}
where $Q_{s}$ denotes the synchrotron tune, $C$ the circumference,
and $\alpha_{\rm C}$ the momentum compaction factor. 

In the case of zero IP dispersion, beamstrahlung excites the
beam particles only longitudinally, and the total energy spread 
follows from the self-consistency relation \cite{ohmifz,ipac16bs} 
\begin{equation}
\sigma_{\delta,{\rm tot}}^{2} = \sigma_{\delta,{\rm SR}}^2 +
\frac{V}{\sigma_{\delta,{\rm tot}}^{2}{\beta^{\ast}_{x}}^{3/2} \varepsilon_{x,{\rm tot}}^{3/2}}\; ,
\label{sz3} 
\end{equation}
where we have introduced the coefficient 
\begin{equation}
V \equiv 
   \frac{25\times 11 }{4\times 36 \pi^{3/2}\alpha} 
\;  Z_{\rm c} \; 
\frac{n_{\rm IP} \tau_{E,{\rm SR}}}{T_{\rm rev}}
  \;  
\frac{r_{e}^{5}N_{b}^{3}\gamma^{2} (2\pi Q_{s})^{2} }{
(\alpha_{\rm C} C)^{2} 
\sqrt{1+\Phi_{\rm piw}^{2}} }\; . 
\end{equation} 
in which the correction factor
(\ref{zcor}) enters.

Table \ref{param} presents example parameters from the 
FCC-ee and CEPC designs. 
The strong impact of beamstrahlung is evident
when comparing the rms bunch length and momentum 
spread due to standard arc synchrotron
radiation, $\sigma_{z {\rm SR}}$  and 
$\sigma_{\delta {\rm SR}}$, and the corresponding
values in collision, 
$\sigma_{z {\rm BS}}$ and 
$\sigma_{\delta {\rm BS}}$.
Beamstrahlung increases the bunch length
and momentum spread by a factor
ranging from about 2 to 4, depending 
on the beam energy.

\begin{table}[htbp]
\caption{Example beam parameters for CEPC Higgs production \cite{cepc}
and three operation modes of FCC-ee \cite{fccee}, illustrating the effect
beamstrahlung on the rms relative momentum spread, $\sigma_{\delta {\rm BS}}$, 
and on the rms bunch length, $\sigma_{z {\rm BS}}$, according to Eqs.~(\ref{sz}) and (\ref{sz3}),
for $n_{\rm IP}=2$ identical IPs. 
The analytically computed
values can be compared with the result of 
beam-beam tracking simulations for FCC-ee, namely the values 
$\sigma_{\delta {\rm SIM}}$ and
$\sigma_{z {\rm SIM}}$) shown underneath \cite{fccee}.  
Parameters calculated in this letter
are shown in bold. 
The last two rows indicate the tolerances on spurious IP dispersion
for a transverse emittance growth of less than 10\%,
based on Eqs.~(\ref{tolx} ) and (\ref{toly}), respectively.
The FCC-ee simulation values (subindex ``SIM'') 
are from D.~Shatilov \cite{fccee}. 
}
\label{param}
\begin{center}
\begin{tabular}{lccccccc}
\hline\hline
Machine  & CEPC & FCC & FCC & FCC  \\
Mode  &  ZH & Z & WW & ZH \\
\hline
beam energy $E_{b}$ [GeV] &120 &45.6 &80 &120 \\
circumference~$C$ [km] &100.02&97.76&97.76&97.76 \\
crossing angle $\theta_{c}$ [mrad] &33&30&30&30\\
bunches/beam $n_{b}$ & 242 & 16640 & 2000 & 328\\
bunch population $N_{b}$ [$10^{10}$] &15 &17 &15 &18 \\
hor.~emittance $\varepsilon_{x}$ [nm] &1.21&0.27&0.84&0.63 \\
vert.~emittance $\varepsilon_{y}$ [pm] &2.40&1.00&1.70&1.30\\
mom.~compaction $\alpha_{\rm C}$ [$10^{-6}$] &11.10&14.80&14.80&7.30\\
hor.~IP beta $\beta_{x }^{\ast}$ [m] &0.36 &0.15&0.20&0.30\\
vert.~IP beta $\beta_{y}^{\ast}$ [mm] &1.5&0.8&1.0&1.0\\
bunch length $\sigma_{z {\rm SR}}$ [mm] & 2.72 &3.50&3.00&3.14\\
{\bf bunch length $\mathbf{\bm{\sigma}_{z {\rm \bf BS}}}$ [mm] } & 
{\bf 3.76} & {\bf 12.58} & {\bf 5.76} & {\bf 5.15}
\\
bunch length $\sigma_{z {\rm SIM}}$ [mm] & --- & 12.1  &  6.0 & 5.3 \\

mom.~spread $\sigma_{\delta {\rm SR}}$ [$\%$] &0.10 &0.038 &0.066 &0.099\\
{\bf 
mom.~spread 
$\mathbf{\bm{\sigma}_{\delta {\rm \bf BS}}}$ [$\mathbf{\%}$]} & {\bf 0.138} & 
{\bf 0.139} & {\bf 0.128} & {\bf 0.165}   \\
mom.~spread 
$\sigma_{\delta {\rm SIM}}$ [$\%$] & --- 
& 0.132 & 0.131 & 0.165   \\
{\bf Piwinski angle $\mathbf{\Phi_{\bf {\rm piw}, BS}}$ } & {\bf 2.97} & {\bf 29.7}  & {\bf 6.7} & {\bf 5.62} \\
energy loss / turn $U_{\rm 0 }$ [GeV] &1.73 &0.036 &0.34&1.72\\
rev.~frequency $f_{\rm rev}$ [Hz] &3003 & 3000 &3000 & 3000\\
RF frequency $f_{\rm RF}$ [MHz] & 650 & 400 & 400 & 400 \\
RF voltage $V_{\rm rf}$ [GV] & 2.17 &0.10 &0.75 &2.0\\
synchrotron tune $Q_{s}$  & 0.065  &0.025 &0.051&0.036\\
longit.~damp.~time $\tau_{E}$ [ms] &23.4 &418.3 &77.5 &23.0\\
rev.~period $T_{rev}$ [ms] &0.33 &0.33&0.33&0.33\\
no.~IPs $n_{IP}$  &2&2&2&2\\
$L_{\rm CDR}$ [$10^{35}$~cm$^{-2}$s$^{-1}$] &0.30&23.00&2.80&0.85\\
{\bf $\mathbf{ \Upsilon_{\rm max}}$ [$\mathbf{10^{-4}}$] } &
{\bf 13.5} & {\bf 14.8} & {\bf 13.1 }& {\bf 21.2}\\
{\bf $\mathbf{ \Upsilon_{\rm ave}}$ [$\mathbf{10^{-4}}$] } & {\bf 5.6} & {\bf 6.1} & {\bf 5.5 } & {\bf 8.9} \\
{\bf 
undulator  parameter 
$\mathbf{K_{\rm max}}$ }& {\bf 6.5} & {\bf 2.6} & {\bf 4.7 } & {\bf 6.4 } \\
{\bf correction factor $\mathbf{Z_{\rm c}}$}  &  {\bf 1.47} & {\bf 2.46} & {\bf 1.70 }
& {\bf 1.69} \\
{\bf $\mathbf{\Delta {\cal H}_{x}^{\ast}}$ [$\mathbf{\bm{\mu}}$m]
($\mathbf{\Delta \bm{\varepsilon}_{x}<0.1\varepsilon_{x}}$)} & {\bf 28.7} & {\bf 0.15} & {\bf 4.94} & {\bf 4.15} \\
{\bf $\mathbf{\Delta {\cal H}_{y}^{\ast}}$ [nm]
($\mathbf{\Delta \bm{\varepsilon}_{y}<0.1\varepsilon_{y}}$)}
& {\bf 56.9} & {\bf 0.58 } & {\bf 10.01} & {\bf 8.57} \\
\hline\hline
\end{tabular}
\end{center}
\end{table}

In the presence of nonzero IP dispersion, also the transverse emittance  
increases due to the beamstrahlung. 
Considering a small spurious horizontal dispersion
at the interaction point (IP), and assuming that   
$|D_{x}^{\ast}|\sigma_{\delta,{\rm tot}}\ll\sqrt{\beta_{x}^{\ast}\varepsilon_{x}}$, 
$\varepsilon_{x,{\rm tot}}$ is no longer constant, but 
determined by the additional equation 
\begin{equation}
\varepsilon_{x,{\rm tot}} \approx \varepsilon_{x,{\rm SR}} + 
\frac{2V {\cal H}_{x}^{\ast}}{\sigma_{\delta,{\rm tot}}^{2}{\beta^{\ast}_{x}}^{3/2} \varepsilon_{x,{\rm tot}}^{3/2}}
\label{ex3} \; , 
\end{equation} 
which needs to be solved self-consistently together with (\ref{sz3}).
The equivalent equation, for the case of spurious vertical 
dispersion, applies to the vertical emittance:
\begin{equation}
\varepsilon_{y,{\rm tot}} \approx \varepsilon_{y,{\rm SR}} + 
\frac{2V {\cal H}_{y}^{\ast}}{\sigma_{\delta,{\rm tot}}^{2}{\beta^{\ast}_{x}}^{3/2} \varepsilon_{x,{\rm tot}}^{3/2}}
\label{ey3} \; . 
\end{equation} 

The spurious dispersion at the IP should not be so large
as to lead to significant emittance blow up $\Delta \epsilon_{x(y)}/\epsilon_{x,(y)}$
From Eqs.~(\ref{ex3}) and (\ref{ey3}) we derive the corresponding 
tolerances for the IP dispersion, namely 
\begin{equation}
|{\cal H}_{x}^{\ast} | <  \frac{\sigma_{\delta,{\rm tot}}^{2}{\beta^{\ast}_{x}}^{3/2} 
\varepsilon_{x,{\rm SR}}^{5/2}}{2 V}
\left[
\frac{\Delta \varepsilon_{x}}{\varepsilon_{x}}
\right]
\label{tolx}
\end{equation}
and
\begin{equation}
|{\cal H}_{y}^{\ast} | <  \frac{\sigma_{\delta,{\rm tot}}^{2}{\beta^{\ast}_{x}}^{3/2} \varepsilon_{x,{\rm SR}}^{3/2}
\varepsilon_{y,{\rm SR}} }{2 V}
\left[
\frac{\Delta \varepsilon_{y}}{\varepsilon_{y}}
\right]
\; . 
\label{toly}
\end{equation}

The resulting tolerances on the two dispersion invariants, 
for a maximum blow up of 10\% are shown in the last two rows
of Table \ref{param}.

In conclusion,
beamstrahlung greatly affects the equilibrium beam distribution in future 
circular Higgs (or electroweak) factories, in particular momentum spread
and bunch length, which must be taken into account when
designing the next generation of lepton colliders, in addition to the constraints
reported in \cite{telnov,Ohmi2017cwi}. 
The beamstrahlung effect also introduces new tolerances on the IP optics parameters.

We thank A.~Blondel, D.~El Khechen, P.~Janot, 
K.~Ohmi, K.~Oide,  
D.~Shatilov, V.~Telnov, and K.~Yokoya 
for helpful discussions. 

\bibliography{beamstrahlung}

\end{document}